\documentclass[fleqn,usenatbib]{mnras}

\usepackage{newtxtext,newtxmath}

\usepackage[T1]{fontenc}

\DeclareRobustCommand{\VAN}[3]{#2}
\let\VANthebibliography\thebibliography
\def\thebibliography{\DeclareRobustCommand{\VAN}[3]{##3}\VANthebibliography}

\usepackage{graphicx}	
\usepackage{amsmath}	
\usepackage{tikz}
\usetikzlibrary{arrows.meta, positioning, shapes, calc}

\title[Fast Multipole Method in RAMSES]{A Scalable Fast Multipole Method Poisson Solver for the RAMSES code: I. Unigrid Algorithm}

\author[J. -Y. Lee and R. Teyssier]{
Jun-Young Lee,$^{1}$\thanks{E-mail: junyoung.lee@princeton.edu}
Romain Teyssier,$^{1}$
\\
$^{1}$Department of Astrophysical Sciences, Princeton University, 4 Ivy Lane, Princeton, NJ 08544, USA\\
}

\date{Accepted XXX. Received YYY; in original form ZZZ}

\pubyear{\the\year{}}

\begin{document}
\label{firstpage}
\pagerange{\pageref{firstpage}--\pageref{lastpage}}
\maketitle

\begin{abstract}
We present a scalable Poisson solver with $\mathcal{O}(N)$ complexity based on the fast multipole method (FMM) implemented in RAMSES. Our FMM constructs a hierarchy of FMM grids on top of the pre-existing Cartesian grid which is used to compute the force for hydrodynamics or particle–mesh simulations. In contrast to the $\mathcal{O}(N)$ multigrid solver (MG) — an iterative method that requires multiple V-cycles through a multi-resolution hierarchy of Cartesian grids — the FMM algorithm performs just one upward pass through the same hierarchy, during which multipole expansions are accumulated and shifted, followed by a single downward pass, in which local expansions are propagated. Numerical tests indicate that FMM attains accuracy comparable to that of MG for smooth potentials and is particularly well-suited for problems with isolated boundary conditions, since it avoids the approximate Dirichlet boundary conditions required by MG schemes. Although in theory FMM requires around 30 times more floating-point operations than MG, its higher arithmetic intensity leads to comparable performance and better scalability relative to MG.
\end{abstract}




\section{Introduction}
Understanding the formation and evolution of structures in our universe involves accurately and, more importantly, efficiently solving the gravitational interaction between bodies. As we explore smaller and smaller scales, our state of the art numerical simulations require ever increasing spatial resolution and computational volume in the description of collisionless N-body systems. Therefore, the scalability of Poisson solvers has become a crucial part of accelerating research in galaxy formation and cosmology. 

While directly aggregating the $1/r^2$ force between bodies yields high fidelity, the algorithm scales with a computational complexity of $\mathcal{O}(N^2)$ given $N$ bodies, challenging experiments with large number of bodies. Multiple algorithms have been introduced to overcome this bottleneck, reaching $\mathcal{O}(N\log N)$ or even  $\mathcal{O}(N)$ while still maintaining high accuracy. One classic way with complexity of $\mathcal{O}(N\log N)$ is to hierarchically organize bodies into an oct-tree, and use expansions of multipoles to approximate force contributions from far away sources \citep{Barnes_Hut_1986}. 

An alternative approach is the Particle–Mesh (PM) method, which solves the Poisson equation by depositing particles onto a mesh, computing the gravitational potential on the grid, and interpolating the resulting forces back to the particles. For bodies deposited in a uniform Cartesian grid (unigrid), the Fast Fourier transform (FFT) of $\mathcal{O}(N\log N)$ is widely adopted \citep{Hockney_Eastwood_1988}. However, for adaptive mesh refinement (AMR), where the grid hierarchy is non-uniform, FFT methods become inefficient, and multigrid \citep[MG;][]{BAKHVALOV1966101, FEDORENKO19621092, Brandt1977MultilevelAS} solvers are commonly used instead, such as in RAMSES \citep{RAMSES_2002, GUILLET_2011} and Athena++ \citep{2023ApJS..266....7T}. MG achieves linear complexity $\mathcal{O}(N)$ by combining iterative relaxation (e.g., Jacobi, Gauss–Seidel, Successive Over-Relaxation) on fine grids with corrections computed on progressively coarser grids through restriction and prolongation operations. Low-frequency error components are efficiently reduced on coarser levels and subsequently prolonged back to finer grids, with only a few V-cycles performed until convergence is achieved.

Another efficient $\mathcal{O}(N)$ algorithm is the Fast Multipole Method \citep[FMM;][]{Greengard_Rokhlin_1987,1997AcNum...6..229G,Cheng_Greengard_Rokhlin_1999}. In comparison to MG, which involves multiple V-cycles, it performs a single upward and downward pass along the FMM tree, an oct-tree used in FMM. Forces are hierarchically approximated using Taylor expansions whose coefficients are computed from the multipole moments accumulated during the upward pass and then propagated downward through the tree. FMM has been proven effective in solving systems of wide range of scales, from stellar dynamics \citep{Dehnen2014_stellar_dynamics} to galactic and cosmological scales by codes such as PKDGRAV3 \citep{PKDGRAV3}, GADGET-4 \citep{GADGET-4} and SWIFT \citep{SWIFT}. Additionally, FMM is also adapted for solving magnetostatic fields \citep{Visscher2010} and vortex fields for fluids \citep{PetFMM}.

Although both MG and FMM scale linearly with problem size, MG is often considered to have a smaller prefactor in practice, as FMM requires frequent matrix–vector operations involving large sets of stored multipole coefficients. A similar consideration explains why many unigrid calculations favor FFT-based solvers over MG for moderate particle numbers. However, with the advent of GPU-accelerated and heterogeneous computing architectures in high-performance computing, algorithms with high arithmetic intensity can be executed very efficiently despite a larger theoretical FLOP count. Therefore, we decided to implement FMM in the RAMSES code \citep{RAMSES_2002} and provide an apples-to-apples comparison of scaling performances to a similar level of accuracy between FMM and MG.

In addition to performance, we stress that FMM has been extensively tested and tuned in direct $N$-body and particle-based solvers, while its use and systematic benchmarking in pure PM  (or Adaptive PM) codes like RAMSES have so far been relatively limited \citep[see e.g.][]{Gnedin2019}. We show that FMM can be particularly useful for systems with vacuum boundary conditions by analyzing the error patterns for isolated gravitational systems. This study serves as a prelude to the implementation of FMM in AMR simulations with adaptive time stepping (Lee and Teyssier 2026 in prep.). To keep this extension native to the AMR refinement rules and adaptive time-stepping framework of RAMSES, we implement an in-house module despite the existence of highly scalable FMM libraries \citep[e.g.,][]{PetFMM, exaFMM}. The sketches of the main ideas of the currently undergoing implementation for AMR will be briefly discussed in Section~\ref{sec:conclusion}.

\section{Fast Multipole Method}
The Fast Multipole Method (FMM) efficiently solves the Poisson equation by hierarchically reusing multipoles and local expansion representations of the gravitational field. The algorithm is primarily composed of two phases along the FMM hierarchy: an upward pass and a downward pass. In the upward pass, multipole moments are first computed either from individual particles or from the leaf cells of the AMR grid (called the particle-to-multipole or P2M step)\footnote{Since our FMM implementation is built on top of a PM scheme, the basic source and sink elements are grid cells onto which particle masses have been already deposited, rather than the particles themselves. Nevertheless, for consistency with the standard FMM terminology, we refer to the corresponding cell-based operations using the conventional particle-based names, such as P2M, L2P, and P2P.}. These multipoles are then successively aggregated (the  multipole-to-multipole or M2M step) from the lowest levels of the hierarchy up to the root. During the subsequent downward pass, the force field is decomposed into far-, intermediate-, and near-field contributions. Intermediate-field interactions are evaluated by translating the multipole expansions of well-separated source cells into local expansions about the target cell via a multipole-to-local (M2L) operation, as specified by the interaction list. The accumulated local expansions, which represent the far-field contribution, are then propagated recursively down the hierarchy from parent to child cells using a local-to-local (L2L step) operation making use of a Taylor expansion. In the final evaluation stage, the local expansions are evaluated at the target particles or cells, while the near-field contributions are obtained by direct pairwise, particle-to-particle summation (P2P step).

Prior to execution of the FMM algorithm, the particles are deposited to the leaf cells of the AMR grid through interpolation schemes like Cloud-in-Cell (CIC) or Triangular Shaped Cloud (TSC). This constitutes a key difference from the classical FMM formulation, which constructs an explicit oct-tree over particles, whereas in our implementation the FMM algorithm is applied directly on the grid cell hierarchy. In the present work, this AMR hierarchy is restricted to a uniform Cartesian grid. On top of the existing grid hierarchy, another hierarchy of FMM grids is added. The level difference of the AMR grid and the finest FMM grid can be set as a parameter ($\Delta \ell$), and the coarsest FMM grid will be the largest grid being able to fill the computational domain.\footnote{RAMSES now supports non-cubical computational domains, and therefore, the coarsest FMM grid does not necessarily cover the full domain.} Moreover, our current implementation of FMM is restricted to isolated (vacuum) boundary conditions.

\begin{figure*}
	\includegraphics[width=\textwidth]{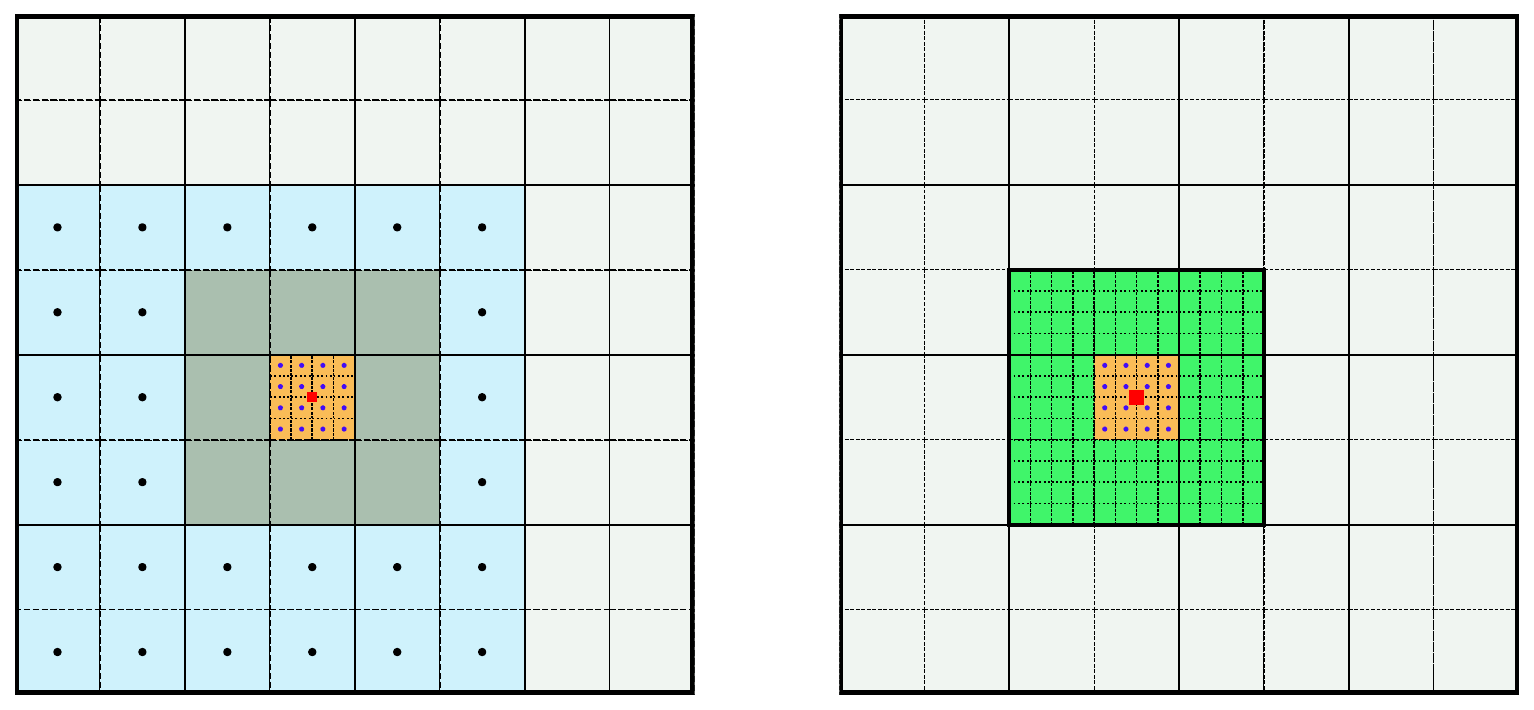}
    \caption{Visualization of the FMM hierarchy used in this paper in two dimensions. We assume a unigrid of $32^2$ cells (AMR level $\ell=5$). The leftmost panel shows a FMM grid of level $\ell=3$, with $8^2$ FMM cells, each with its own multipole and local expansion terms. The orange cell represents the target FMM cell, the dark green cells indicate the near-field region to be accumulated at the next level, and the light blue cells show the interaction list used to calculate the intermediate-field. The red square at the midpoint of the target grid represents the source of the parent FMM cell’s local expansion, which is shifted toward the midpoints of the target FMM cells in blue to account for the far-field. The intermediate-field is calculated using the multipole-to-local (M2L) translation of the FMM cells (\textit{black circles}) in the interaction list. The near-field is resolved at the next finer level, following the same field decomposition scheme in the right panel. Here, the target FMM grid, which is one of the target FMM cells from the previous level, contains sixteen AMR cells. In other words, the level difference between the finest FMM grid and this grid is in this example $\Delta \ell=2$. Once the downward pass reaches the bottom of the hierarchy, the near-field is calculated via direct force summation, as shown in the right panel. For each fine cell, all gravitational contributions from cells in the colored region are computed using the Green’s function $g(r)$, with $r$ taken as the cell displacement. For the cell itself, $r$ is set to a minimum of the physical cell size, $\Delta x$.}
    \label{fig:interaction}
\end{figure*}

In the following subsections, we describe the algorithmic components in detail. Section \ref{sec:upward} presents the upward pass, including the P2M and M2M operations. Section \ref{sec:interaction} then introduces the interaction list, which partitions source–target interactions into far-, intermediate-, and near-field contributions. This decomposition is employed in the downward pass, where the M2L, L2L, and P2P operations are carried out, as described in Section \ref{sec:downward}. Our formulation largely follows \citet{Dehnen2002_theory}, adopting outer-product representations and Einstein summation conventions instead of multi-index notation. 

\subsection{Upward Pass: Multipole Accumulation}
\label{sec:upward}
The integral step common to many gravity solvers involves a coarse representation of the density field through a hierarchy of multipoles. The multipoles are calculated from the leaf cells in the following discretized form:
\begin{eqnarray} \label{eqn:multipole}
    \mathcal{M}_n = \sum_{\rm cell_i \in grid} m_{\rm cell} \,  {\bf x_i}^{(n)},
\end{eqnarray}
where $\mathcal{M}_n$ is the $n$th multipole, $m_i$ the mass in the cell, and ${\bf x_i}$ the position of the cell-center with respect to the origin. The superscript of $(n)$ denotes the $n$-fold outer product. 

In our implementation, we retain terms up to the quadrupoles ($n=2$). Using higher-order multipoles would require increasingly larger memory, as it will require $\binom{d+n-1}{n}$ for an $n$th order multipole, or cumulatively, $\sum_{k=0}^{n} \binom{d+k-1}{k} = \binom{d+n}{n}$, in $d$-dimensions. With $d=3$ and $n=2$, we need to store only 10 elements per FMM cell. Extending the expansion to include octopoles could improve accuracy but would double the memory cost. This could be alleviated using a shallower FMM grid (in our formulation a larger level difference $\Delta \ell$) but at the expense of a more costly P2P steps (more later). Our accuracy tests in Section \ref{sec:test} show that such a low polynomial order is sufficient enough to be comparable with MG. As our paper aims at comparing MG and FMM solvers at controlled accuracy in the PM code, we design our implementation to be limited to quadrupoles, following \cite{Dehnen2002_theory}. Comparing the two methods at higher polynomial order would also be an interesting exercise, requiring both a higher-order finite-difference approximation to the Laplacian operator for MG and higher-order multipole expansions for FMM; however, this is beyond the scope of the present paper.

Another possibility of reducing the memory and computational cost in $d=3$ is by expanding the multipoles in spherical harmonics rather than in Cartesian coordinates, since this approach requires only $(n+1)^2$ elements \citep{1997AcNum...6..229G}. Nevertheless, for the expansion order considered here, $n=2$, the improvement is marginal and therefore we retain the Cartesian expansion scheme, similar to \cite{Visscher2010}. Our accuracy and scalability tests (Sections~\ref{sec:test} and \ref{sec:scaling}) show that this implementation strikes a good balance between computational efficiency and solution accuracy, with performance comparable to the MG solver in RAMSES. Since our primary target applications do not require the extremely high precision associated with direct $N$-body simulations in stellar dynamics, we focus on achieving robust and scalable performance within the accuracy range relevant to our problems.

Once the multipole representation has been computed from the leaf cells (P2M), it is subsequently accumulated in a bottom-up fashion (M2M). The accumulation is first carried out by directly adding the masses of the leaf cells to the finest FMM cells. In the RAMSES code, the basic computational element is an {\it oct}, a small grid holding 8 cells. Each FMM oct will thus hold 80 multipole coefficients. Moreover, for a level offset $\Delta \ell$ between the coarsest AMR grid and the finest FMM grid, and assuming a unigrid configuration in which all AMR cells are leaf cells at the deepest level, the multipole moments from $8^{\Delta \ell}$ cells are merged into a single parent cell in the FMM hierarchy. Since all multipoles are defined with respect to the origin ($\mathbf{x}=0$ in the box frame), this merging operation simplifies to a direct term-by-term summation.

Finally, we traverse the FMM grid hierarchy bottom-up once again to shift the multipoles from the origin to the center of the FMM cell $\mathbf{a}$ via the following equation: 
\begin{eqnarray}
\label{eq:M2M_general}
\mathcal{M}'_n =
\sum_{k=0}^{n}
\binom{n}{k}
(-\mathbf{a})^{(n-k)}
\odot
\mathcal{M}_k,
\end{eqnarray}
where $\odot$ is the tensor inner product. Using cell-centered multipoles improves computational efficiency by keeping the interaction geometry fixed, whereas centering the expansion on the center of mass would change the geometry of the interaction list (see Section \ref{sec:interaction} for more information). The downside is the appearance of a non-zero dipole term and an increase in memory usage, but we opt for the cell-centered scheme in order to take advantage of the speedup offered by pre-computing coefficients. Moreover, although the multipole shifts could be applied during accumulation and folded into a single upward pass, the downward pass is already significantly more expensive than the upward pass, so the potential performance gain is limited. Since most of the accumulation is performed locally and does not involve MPI communication, we keep the accumulation and shifting steps separate.

\subsection{Interaction List Construction}
\label{sec:interaction}

In FMM, the gravitational field is decomposed into a far-field,  an intermediate-field, and a near-field with respect to a given target cell of interest. Forces from different fields treated with different approximation schemes to reduce computational cost while maintaining accuracy. In particular, the interaction list is defined to represent the intermediate-field, which is approximated using multipole expansion terms at the current FMM level. The far- and near-fields correspond to the interaction lists at coarser and finer levels, respectively, where their contributions are evaluated at those associated levels.

Fig.~\ref{fig:interaction} shows the field decomposition and interaction list. Given a target FMM oct, the $2^d$ FMM cells that make up the grid share the same interaction list. The $3^d - 1$ FMM grids at the same level that are directly in contact with the target grid form the near-field, whose contributions are accumulated further down the hierarchy. The $6^d - 3^d$ FMM grids, ignoring the isolated boundary, correspond to the intermediate-field and are resolved using M2L translations. Finally, the far-field consists of the remaining grids in the computational domain, whose contributions have already been computed at previous levels and stored in the parent’s local expansion. These contributions are then shifted to the FMM cell center via the L2L operation, which is described in Section \ref{sec:downward}. As the downward pass reaches the final level of the FMM hierarchy, the near-field is no longer delegated to the next level but calculated via direct summation.

The advantage of using the cell-centered scheme arises in that the interaction list has a fixed geometry. Instead of evaluating the displacements of the center of mass for every individual cell, the displacement can be precomputed in constant time and spatial complexity. In addition, we can precompute the Green’s function along with its Taylor expansion, thus reducing the frequency of costly operations such as divisions. As a tradeoff of this static scheme, an additional cost of evaluating the dipole per cell arises. However, this decision was taken because our implementation goal is to increase the number of FLOPs and reduce conditional branching for future application in GPUs.

Alternative schemes defining interaction lists and well-separated cells are available, based on opening angles similar to the force evaluation criteria of \cite{Barnes_Hut_1986}. Although the original FMM \citep{Greengard_Rokhlin_1987} achieves the target accuracy by adjusting the expansion order, the latter adjusts the opening angle as the tolerance parameter for force errors even in a fixed expansion order \citep{Dehnen2000_symmetry,Dehnen2002_theory}. Although many astrophysical codes have opted for adaptive opening angle criteria \citep{Dehnen2014_stellar_dynamics,PKDGRAV3,GADGET-4,SWIFT}, we deliberately restrict ourselves to rigid interaction geometries. Considering the forthcoming implementation of FMM in the AMR regime, the AMR geometry will naturally introduce spatial adaptivity. More refined cells will be treated at finer interaction scales than the coarser cells. This design reflects the nature of the PM scheme in RAMSES, and we choose not to add an additional adaptive opening angle criterion on top of the refined grid. Moreover, although the number of interactions can be further reduced by choosing an adaptive opening angle, we fix the interaction geometry so that we can take advantage of precomputed translation kernels. Considering a future implementation of the GPU version, this will reduce the necessity of extra conditional checks for multipole acceptance, thereby possibly improving scalability.

\begin{figure*}
	\includegraphics[width=\textwidth]{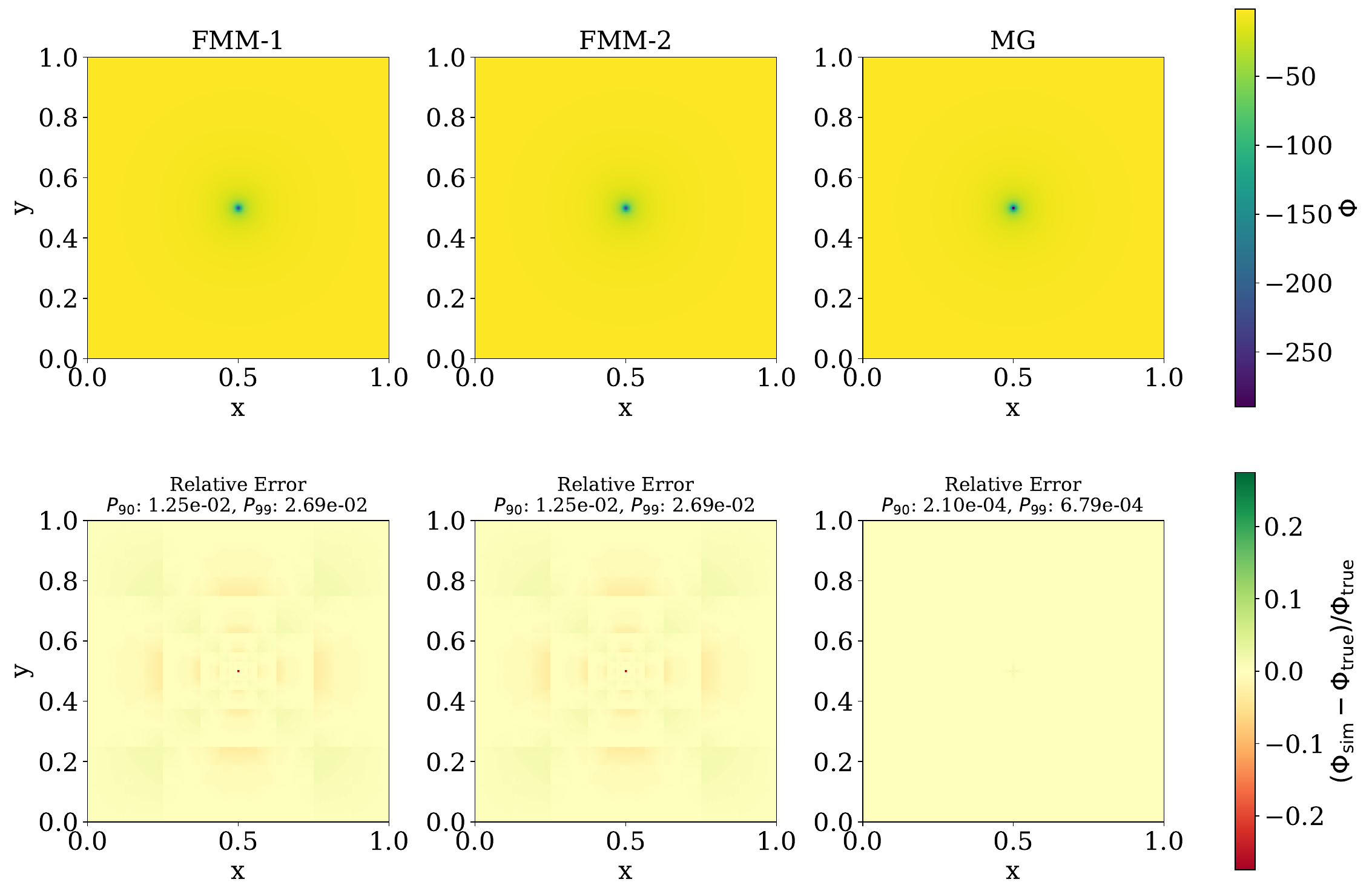}
    \caption{Potential maps (\textit{top}) and relative error maps (\textit{bottom}) for a slice through the plane containing a single point charge with mass $GM=1$. The computational domain has box size 1 and is resolved on a $256^3$ grid. Columns from left to right show results for FMM-1, FMM-2, and MG. For the FMM cases, the number indicates the level difference between the finest AMR cells and the finest FMM cells. We report the 90th-percentile ($P_{90}$) and 99th-percentile ($P_{99}$) of the absolute relative error for each solutions.}
    \label{fig:single}
\end{figure*}

\begin{figure}
	\includegraphics[width=\columnwidth]{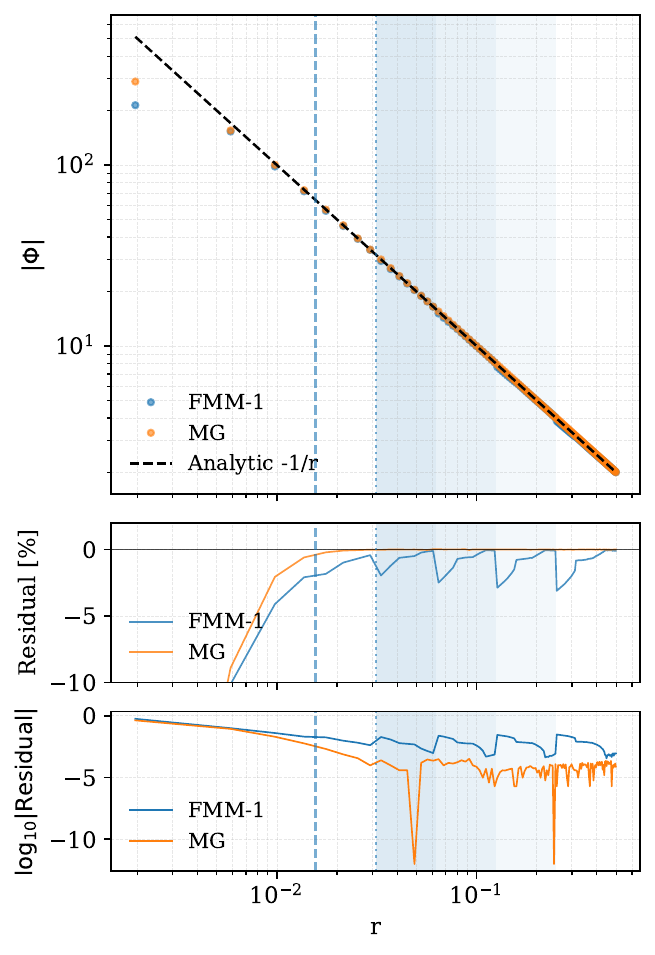}
    \caption{Gravitational potential (\textit{top}) for the single-charge configuration. Results from FMM-1 (\textit{blue}), MG (\textit{orange}), and the analytic solution (\textit{black dashed line}) are shown. The \textit{middle panel} shows the residual, while the bottom panel shows $\log_{10}$ of the residual. Only the $+x$ direction is depicted, corresponding to a slice through the midplane containing the charge. \textit{Blue vertical dashed line} indicates where the charge contributes to the field via direct force summation, while the \textit{blue vertical dotted line} marks the boundary within which the approximations from L2L translations are not used. \textit{Shaded regions} in different colors denote the distinct zones computed using the different levels of local expansion.}
    \label{fig:single_err}
\end{figure}

\begin{figure*}
	\includegraphics[width=\textwidth]{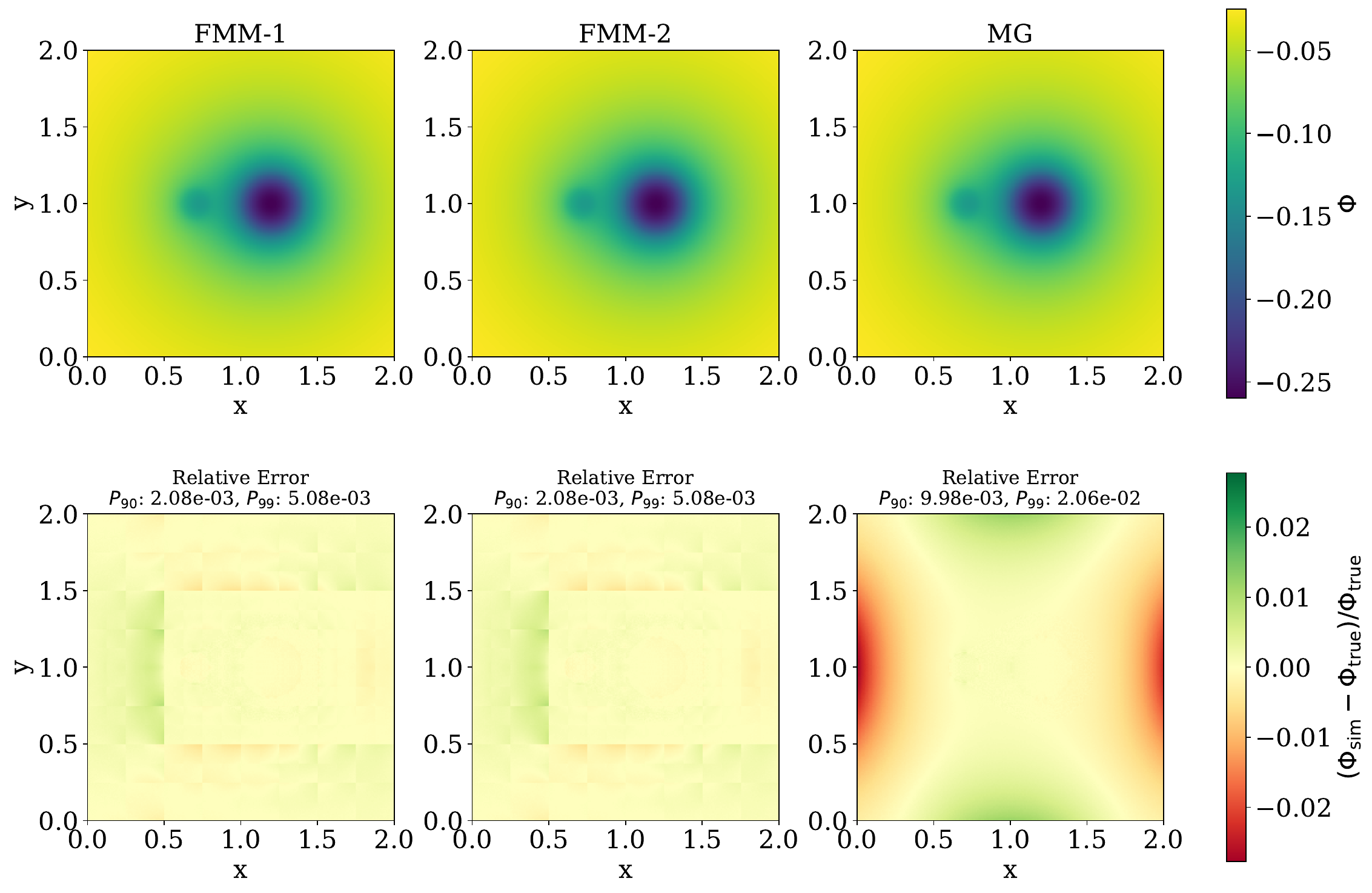}
    \caption{Same as Fig.~\ref{fig:single}, but for two spheres located at $(0.7, 1.0, 1.0)$ and $(1.2, 1.0, 1.0)$ with radii 0.1 and 0.2, respectively, in a box of size 2. The domain is resolved on a $256^3$ grid. Shown is a slice through the plane at $z=1$.}
    \label{fig:spheres}
\end{figure*}

\subsection{\label{sec:downward}Downward Pass: Local Expansion and Direct Summation}

Given the interaction lists defined in Section \ref{sec:interaction} and Fig.~\ref{fig:interaction}, we perform the downward pass, which involves the M2L, L2L and P2P operations. The M2L operation, or multipole-to-local translation, evaluates the contribution of intermediate-field source cells by translating their multipole expansions into a local expansion about the target cell. This is given by
\begin{eqnarray}
\label{eq:M2L}
\mathcal{L}^{p}_m
=
\sum_{n=0}^{p-m}
\frac{(-1)^n}{n!}
\nabla^{(n+m)} g(\mathbf{R})
\odot
\mathcal{M}_n ,
\end{eqnarray}
where $\mathcal{L}^{p}_m$ denotes the $m$-th order local expansion coefficient for a given truncation order $p$, $\mathbf{R} = \mathbf{x}_{\rm source} - \mathbf{x}_{\rm target}$ is the displacement between the source and target cell centers,  and $g(\cdot)$ is the real-space Green’s function. We retain expansion terms up to third order ($p=3$), corresponding to 20 local expansion coefficients per FMM cell, while the octopole contributions are omitted.

In the next level, the intermediate-field becomes the far-field of the child FMM cell. This far-field contribution is assessed at the center of the parent FMM cell, and therefore needs to be corrected to the child's cell center. This local-to-local (L2L) translation is performed simply by a Taylor expansion, 
\begin{eqnarray}
\label{eq:L2L}
{\mathcal{L}^{p}_m}^\prime
=
\sum_{n=0}^{p-m}
\frac{1}{n!}
\mathbf{a}^{(n)}
\odot
\mathcal{L}^{p}_{m+n} ,
\end{eqnarray}
where the unprimed tensor is evaluated in the center of the parent cell, the primed tensor in the center of the child cell, and $\mathbf{a}$ is the displacement vector from parent to child FMM cell. At the finest FMM level, $\mathbf{a}$ becomes the displacement between the parent FMM cell and the cells of the uniform grid. Each FMM cell will contain $2^{d\Delta \ell}$ of these fine cells, and a separate displacement matrix should be constructed. 

The direct summation (P2P) at the finest level of the hierarchy is computed using the three-dimensional Green's function
\begin{eqnarray}
\label{eq:direct}
g(\mathbf{r}) =
\begin{cases}
-\frac{1}{|\mathbf{r}|}, & |\mathbf{r}| > \Delta x,\\[2mm]
-\frac{1}{\Delta x}, & |\mathbf{r}| \le \Delta x,
\end{cases}
\end{eqnarray}
where $\Delta x$ is the size of the cell (or $\Delta x = L/2^\ell$ at level $\ell$ with the length of the box $\ell$), 
and $\mathbf{r}$ is the displacement between cells. The softening of Green's function at $1/\Delta x$ is a design choice that can be adjusted. However, the exact solution depends on the subgrid density profile, and the results are generally insensitive as long as the resolution is adequate. The near-field contribution,
$\Phi_{\rm near} = \sum_{\rm cell_i \in \text{near}} m_{\rm cell_i} \, g(\mathbf{r}_i)$,
is then added to the previously computed $\Phi_{\rm far+mid}$ to obtain the complete potential field.

It is worth mentioning that FFT can be used in conjunction with FMM to accelerate the computations in different ways.  First, \cite{Gnedin2019} replaces multipole expansions with effective masses assigned to subdivided cells and uses FFT to perform the M2L step. This approach can significantly improve performance relative to standard FMM implementations, while retaining high accuracy using approximately optimal Green’s functions \citep{Gnedin2021}.

Second,  in case of periodic boundary conditions, \cite{Wang2021} adopts a Tree-PM-like approach \citep{TreePM}, decomposing the gravitational force into long- and short-range components, with a PM solver for the former and FMM for the latter, thereby reducing the cost of expensive M2P and P2P operations. Although this approach of FFT and FMM can still be applicable and designed to work efficiently on non-uniform grids, \cite{Gholami2016} has demonstrated that for highly localized fields, FFT itself can be outperformed by FMM and MG. This is also true for very large base Cartesian grid. Moreover, as our primary goal is to develop a solver that operates directly on the AMR grid and is naturally coupled to the hydrodynamics solver, we adopt a pure FMM formulation and don't consider any FFT component in our implementation.

\begin{figure}
	\includegraphics[width=\columnwidth]{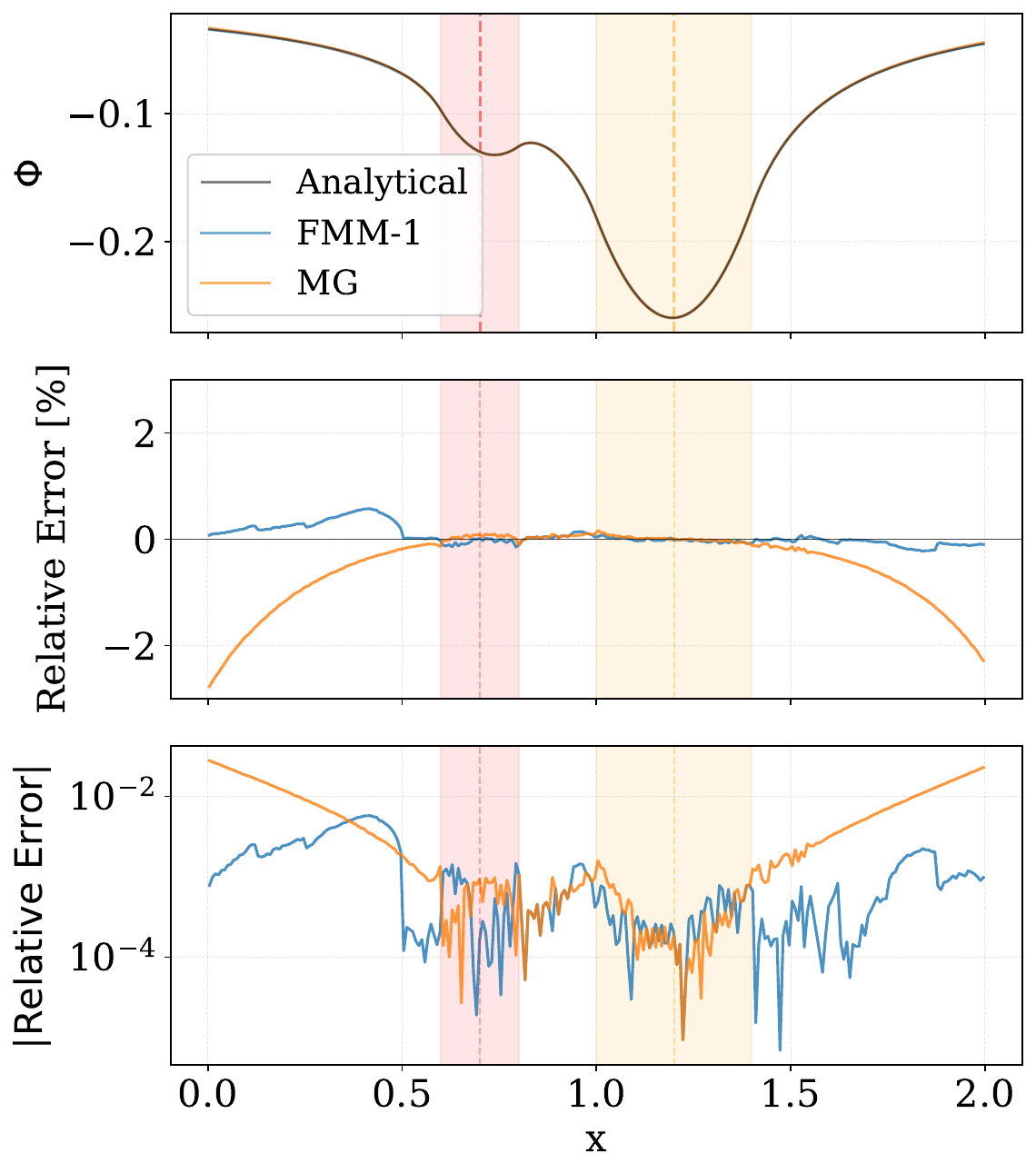}
    \caption{Same as Fig.~\ref{fig:single_err}, but for the double-sphere configuration. Results for FMM-1 are plotted in \textit{blue} and for MG in \textit{orange solid lines}. The \textit{red} and \textit{yellow shaded areas} indicate the interior of each sphere, and the \textit{vertical dashed lines} mark their centers. Overall, FMM-1 produces more accurate solutions than MG, with particularly improved accuracy near the boundaries. However, we emphasize that the solutions influencing the system’s dynamics are restricted to the inner regions of the sphere, where mass is present. Therefore, any bias introduced by the boundary conditions for MG, or the boxy errors outside the spheres for FMM, is likely not physically relevant.}
    \label{fig:spheres_err}
\end{figure}

\section{Test Results}
\label{sec:test}
In this section, we present test results for simple systems in a three-dimensional, isolated box. We compare the potential field map of a given system from MG\footnote{For the MG solver, we use two Gauss–Seidel pre- and post-smoothing iterations and adopt a tolerance (stopping criterion) of $10^{-6}$, which are standard choices in MG in the RAMSES code \citep{GUILLET_2011}. This tolerance is also kept fixed for the subsequent scaling experiments in Section \ref{sec:scaling}, where we observe that the resulting solution accuracies remain comparable with FMM. This tolerance usually corresponds to 6 V-cycles. } and FMM methods, where FMM-1 denotes a level difference of $\Delta \ell=1$ and FMM-2, $\Delta \ell=2$. All test particles are deposited onto the grid using the CIC method.\footnote{Although it is natural for FMM-2 to employ higher-order multipoles than FMM-1, our subsequent scaling results show that it already scales worse than MG. We therefore restrict ourselves to the minimal modification, fixing the expansion at quadrupole order.} We first evaluate our method in the simplest configuration: a single charge located at the center of the box (Fig.~\ref{fig:single}). Then we test a smoother density configuration with two uniform-density spheres in Fig.~\ref{fig:spheres}. We provide an additional astrophysical setup with a single isolated NFW halo in Appendix~\ref{sec:nfw}.

We first evaluate our method in the simplest configuration: a single charge located at the center of the box. Fig.~\ref{fig:single} presents the potential field map and the corresponding error distribution along a midplane slice that contains a single point mass with $GM=1$. We observe that the 90th- and 99th-percentile errors for both FMM-1 and FMM-2 is approximately two orders of magnitude larger than for MG. In addition, the quality of the solution is nearly indistinguishable between FMM-1 and FMM-2. The error maps for FMM exhibit a characteristic boxy, highly-correlated, and symmetric structure, which arises from discontinuities in the local expansions across cell boundaries. Fig.~\ref{fig:single_err} presents the solution error for FMM-1 and MG, plotted against the analytical solution along the $+x$ direction. Once more, the MG method yields higher-quality solutions, and the observed spikes in the FMM solution occur precisely at the interfaces where local expansions of different levels are used, which accounts for the boxy structures visible in the error map.\footnote{While boxy error patterns are an intrinsic limitation of the FMM solver, they can be alleviated by applying random affine transformations to the computational domain \citep{Barnes2026}. Additionally, multigrid (MG) solutions, which are usually smooth, can deteriorate on AMR grids compared to uniform grids, especially near the interfaces between different refinement levels.} We note that although the error values for a single charge case is severely worse for FMM, we find that we achieve a similar or slightly better level of accuracy for extended density distributions such as two uniform density spheres (see Fig.~\ref{fig:spheres} and Fig.~\ref{fig:spheres_err}) and an NFW profile (see Fig.~\ref{fig:nfw} and Fig.~\ref{fig:nfw_err}).

Lastly, we observe the expected decrease in the absolute value of the potential for both MG and FMM-1 near the source. This decrease is more pronounced for FMM-1 and is physically accurate, since the nearest cell is computed solely by direct force summation. The drop is an inherent consequence of the CIC mass deposition scheme, which spreads the mass of the single charge over eight neighboring cells.

Since the FMM method takes advantage of the local expansion, it struggles for discrete density fields such as the single charge configuration. We thus test for a smoother density field with two uniform-density spheres as in \cite{2023ApJS..266....7T}. In a box of size 2, resolved with a grid of $256^3$ cells, we locate two spheres at $(0.7, 1.0, 1.0)$ and $(1.2, 1.0, 1.0)$ with radii 0.1 and 0.2, respectively. Fig.~\ref{fig:spheres} presents the potential maps and corresponding error maps for the FMM-1, FMM-2, and MG schemes. 

As before, FMM-1 and FMM-2 produce nearly indistinguishable results, characterized by the same boxy error patterns. In contrast to the single-charge configuration, however, the MG approach performs noticeably worse, particularly near the box boundaries, where both the 90th- and 99th-percentile errors increase by a few factors. This degradation comes from the approximate Dirichlet boundary conditions employed in the MG solver. Fig.~\ref{fig:spheres_err} illustrates this behavior more clearly by plotting the error along the $x$-axis. Within the spheres, where the self-induced gravitational potential is strictly parabolic, the field is now accurately captured by the FMM approach, unlike in the single-charge case. Although the MG method achieves comparable accuracy inside the spheres, its errors increase near the boundaries, so the FMM solvers provide substantially more reliable results. Improving the accuracy of the Dirichlet boundary conditions, using, for example, the \cite{James1977} method, can reduce the magnitude of the error for MG, at the expense of running the MG solver twice. 

\begin{figure*}
	\includegraphics[width=\textwidth]{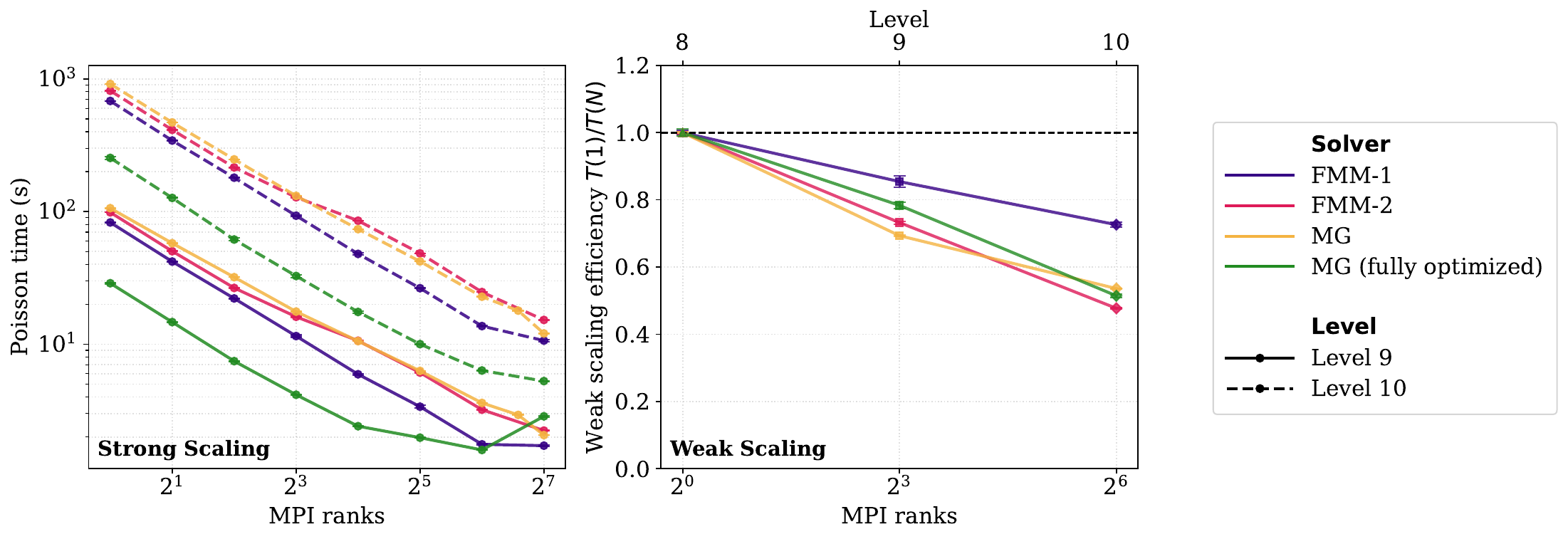}
    \caption{Strong scaling (\textit{left}) and weak scaling (\textit{right}) results comparing the FMM-1 (\textit{purple}), FMM-2 (\textit{pink}), MG in RAMSES (\textit{yellow}), and fully-optimized MG in RAMSES (\textit{green}). To ensure statistical stability, each data point is obtained from 10 independent trials, with error bars indicating the $1\sigma$ standard deviation. Each trial averages over five time steps. We report the wall time spent exclusively within the solver routines. The strong scaling analysis examines performance as a function of the number of MPI ranks for fixed problem sizes at levels 9 (\textit{solid lines}) and 10 (\textit{dashed lines}). For the weak scaling analysis, the workload per MPI rank is held fixed while the total number of MPI ranks increases by a factor of eight per level. We report the parallel efficiency normalized as $T(1)/T(N)$, where $T(1)$ is the runtime on a single MPI rank at level 8.}
    \label{fig:scaling}
\end{figure*}

\section{Performance Scaling}
\label{sec:scaling}
In this section, we compare the scaling behavior of the FMM solver with that of the MG solver. Although both solvers exhibit $\mathcal{O}(N)$ scaling, the FMM method is known to have a significantly larger prefactor, which increases practical computational cost. However, the FMM benefits from the fact that it needs less frequent MPI communications, as it performs a single sweep upward and downward of the FMM hierarchy, unlike MG, which needs multiple iterations for convergence. For MG, we present two versions, one with our recent implementation\footnote{We use the prototype code {\ttfamily mini-ramses} to develop our FMM scheme and compare to the MG version there.} similar to the current FMM algorithm, and the other with full optimization. The fully optimized MG\footnote{The fully optimized MG solver is available in the public repository of the original RAMSES code.} overcomes MPI communication bottlenecks via calling bulks of neighboring cell information and increases efficiency via vectorization. However, as a trade-off, the optimized version often requires a memory footprint more than twice as large as that of the non-optimized solver.

We perform both strong and weak scaling tests using the NFW halo initial condition at multiple levels.  For the strong scaling tests, we consider levels 9 and 10, corresponding to resolutions of $256^3$ and $512^3$ cells,\footnote{The resolution is given by $2^{3(\ell -1)}$ for level $\ell$, since the coarsest level is used to construct the computational domain with isolated boundary conditions.} and vary the number of MPI ranks as $2^n$, with $n = 0, \ldots, 7$. For the weak scaling tests, we keep the workload per MPI rank approximately fixed by solving the problem at level 8 with one rank, level 9 with 8 ranks, and level 10 with 64 ranks\footnote{Our weak scaling setup differs slightly from the conventional setup, in which the domain is tiled by a factor of $N$ along each spatial dimension at fixed hierarchy depth. Here, we instead keep the number of cells per MPI rank approximately fixed while increasing the total problem size by deepening the FMM hierarchy. This makes the test more stringent for the solvers, since $\mathcal{O}(N\log N)$ algorithms incur an additional logarithmic cost as the depth increases, whereas algorithms with linear complexity should be less affected.}. We report the efficiency by $T(1)/T(N)$, where $T(1)$ is the runtime for the a single rank solving for level 8. Each solver is evolved for five fixed time steps, and each configuration is run 10 times. For each data point, we report the mean wall-clock time spent exclusively in the solvers per time step, together with the corresponding $1\sigma$ standard deviations.

Interestingly, from the strong scaling results, we find that all three solvers behave very similar up to four ranks except for the fully optimized MG. FMM-1 always scales better than MG and FMM-2, with a more robust behavior as it almost scales as a power-law. Given that the accuracy gain from using FMM-2 compared to FMM-1 is negligible, we find that FMM-1 is the most optimal option in most applications. The competitive advantage of FMM-1 increases as we add more MPI ranks, until we reach strong scaling saturation at 128 ranks for FMM-1. 

For weak scaling experiments, we find that FMM-1 outperforms even the fully optimized MG implementation. Although the efficiency decreases for all solvers as both the problem size and the rank increase, the FMM-1 solver still scales better than the MG solver. Especially, FMM-1 demonstrates superior scaling compared to FMM-2, further confirming that FMM-1 is the preferable option across a wide range of scenarios.

The increased runtime of FMM-2 relative to FMM-1 can be explained by the fact that direct summation accounts for a larger fraction of the overall computation. To illustrate this, consider a finest grid at level $\ell$, with $\Delta \ell$ again denoting the level offset between the finest FMM grid and the unigrid. Since the upward pass consists primarily of local operations and is substantially faster than the downward pass, we restrict our attention to the latter. We decompose the computational cost of the downward pass into contributions from the far-field L2L and L2P translation ($\mathcal{F}_{\ell'}$), the intermediate-field M2L and M2P translation ($\mathcal{M}_{\ell'}$) and the direct summation in the near-field of the finest level ($\mathcal{D}_{\ell'}$). The terms of intermediate-field and direct-summation are indexed by $\ell'$ because, for a fixed number of MPI ranks, the MPI communication will be more frequent as fewer octs are assigned per rank. By contrast, L2L translations are performed almost entirely locally, except at the coarsest levels, where computational work is minimal. 

Consider we are solving the Poisson equation for level $\ell$ with $8^\ell$ cells, and the FMM grid lies above with a level offset of $\Delta \ell$. Per cell, the multipole accumulation is solved in $\mathcal{U}$ operations, far-field is solved in $\mathcal{F}_{\ell'}$ operations, the intermediate-field in $(6^3-3^3)\mathcal{M}_{\ell -\Delta \ell+1}$ operations, and the near-field in $(27\times 8^{\Delta \ell})\mathcal{D}_{\ell -\Delta \ell+1}$ operations. 

Now, considering levels above starting from base FMM level of $\ell'=1$, and ignoring the boundary effects, the total time accumulated over all MPI ranks is
\begin{align} \label{eqn:Tfmm_complex}
T_{\rm FMM} (\ell ;\Delta \ell) &\approx
\Bigl(\sum_{\ell '=1}^{\ell -\Delta \ell} 8^{\ell'} + 8^\ell  \Bigr) \mathcal{U}
+ 189 \Bigl(\sum_{\ell '=1}^{\ell -\Delta \ell} 8^{\ell '} \mathcal{M}_{\ell '} + 8^{\ell}\mathcal{M}_{\ell -\Delta \ell + 1}\Bigr) \nonumber\\
&\quad + \Bigl(\sum_{\ell '=1}^{\ell -\Delta \ell} 8^{\ell '}\mathcal{F}_{\ell  '} + 8^\ell \mathcal{F}_{\ell -\Delta \ell + 1} \Bigr)
+ 8^\ell \Bigl(27 \times 8^{\Delta \ell} \Bigr)
\mathcal{D}_{\ell - \Delta \ell + 1}.
\end{align}
Now comparing,  $\Delta T \equiv T_{\rm FMM} (\ell ;\Delta \ell=2) - T_{\rm FMM} (\ell ;\Delta \ell=1)$, 
\begin{eqnarray}
\Delta T \approx 8^\ell\Bigg[
189\left(\frac{7}{8}\mathcal{M}_{\ell -1} - \mathcal{M}_{\ell}\right)
+ 1728\,\mathcal{D}_{\ell -1}
- 216\,\mathcal{D}_{\ell}
\Bigg],
\end{eqnarray}
where we neglect the subdominant multipole accumulation and the far field contributions.

Neglecting the subdominant far-field contribution, the difference reflects a competition between increased near-field and decreased intermediate-field computations. On a single core, the level-dependent MPI communication overhead can be neglected, i.e., $\mathcal{O}_{\ell -1} \approx O_\ell$, which reduces the expression for the difference to $\Delta T \approx 8^\ell \left[ -\tfrac{189}{8} \mathcal{M}_\ell + 1512 \mathcal{D}_\ell \right]$. Although the prefactor is smaller for the intermediate-field contribution, the operation count is significantly larger compared to the direct summation, as it utilizes M2P expansions from 10 multipole components. We find that, on a single core, the reduction in P2P interactions offset by the increased use of M2P operations leads to only a modest performance advantage of FMM-1 over FMM-2.

We now consider how these level-dependent operations behave as a function of the number of MPI ranks. The main scaling benefit of FMM-1 over FMM-2 likewise arises from FMM-2’s greater dependence on P2P operations, which exhibit lower arithmetic intensity and are thus more susceptible to MPI communication overhead. We also note a small bump indicating some deterioration in the strong scaling performance of FMM-2. Because FMM-2’s near-field region is eight times thicker than that of FMM-1, it triggers a larger volume of MPI communications. A comparable bump is observed for MG, which can similarly be attributed to MPI communication overhead. However, in this case it is caused by many calls originating from thin layers, rather than from a single thick layer as in FMM-2. This explains why the performance gain from using FMM-1 instead of FMM-2 is non-monotonic, rising and then falling, as seen in Fig. \ref{fig:scaling}. 

Counting the number of FLOPs for the FMM operations, the FLOPs for the far-field and intermediate-field differ at the final level, as we will only need the scalar component of the local expansion, $\mathcal{L}^p_0$, for the final evaluation of the potential field. Thus, $\mathcal{U}=10$ (M2M), $\mathcal F_{\ell'<\ell -\Delta \ell+1}=357$ (L2L), $\mathcal F_{\ell -\Delta \ell+1}=74$ (L2P), $\mathcal M_{\ell'<\ell -\Delta \ell+1}=228$ (M2L) and $\mathcal M_{\ell -\Delta \ell+1}=36$ (M2P). The direct summation is always fixed to $\mathcal{D}_{\ell'}=2$ (P2P). Naively calculating the FLOPs would yield about a 20\% higher number of operations for FMM-1 than FMM-2. However, the arithmetic intensity of the direct summation is much lower than that of the intermediate-field operator. As a result, in the memory-bound regime, the compute time of the intermediate-field operation can be comparable to or even smaller than that of the direct summation, making FMM-1 more favorable as seen from the tests.

Before comparing FMM with the MG solver, we simplify the expression in Equation~\ref{eqn:Tfmm_complex}. We ignore the computations above the finest level, simplify all operations so that $\mathcal{U} \sim p^3$ (P2M), $\mathcal{F}\sim p^3$ (L2P), $\mathcal{M} \sim p^3$ (M2P), and $\mathcal{D} \sim 1$ (P2P), where $p$ is the truncation order of the expansions. Moreover, putting $N=8^\ell$ and $s=8^{\Delta \ell}$, the expression reduces to:
\begin{eqnarray}  \label{eqn:Tfmm_reduced}
    T_{\rm FMM} \approx 189N p^3 + 2Np^3 + 27Ns.
\end{eqnarray}
This is similar to the compact form used by \cite{1997AcNum...6..229G}, $T_{\rm FMM} \approx 189 \frac{N}{s} p^4 + 2Np^2 + 27Ns$, 
where $N$ is the total number of particles (total number of cells in our case) and $s$ is the average number of particles per finest FMM cell ($8^{\Delta \ell}$ cells of the unigrid in our case). The key difference lies in the first term: our formulation yields $189 N p^3$, replacing $189 (N/s) p^4$. This occurs because, to achieve higher accuracy at the finest level, we evaluate M2P directly at the cell centers instead of first aggregating them at the FMM cell centers, which would involve M2L operations that cost $\mathcal{O}(p^4)$ for $N/s$ FMM cells (see Fig. \ref{fig:interaction}). Thus, the factor $N/s$ is replaced by $N$, while the dependence on $p$ is reduced by one power. Furthermore, in the second term we have an extra order for $p$, as we use a Cartesian tensor basis instead of spherical harmonics.

The MG solver in RAMSES uses multiple V-cycles to get an accurate solution below the error tolerance, and let the number of cycles be $N_{\rm cycle}$. Each V-cycle is consisted of pre- and post-smoothing with Gauss Seidel smoothing operations ($\mathcal{O}_{\rm GS}$), which can be set as a parameter, and let the total smoothing done for each level be $N_{\rm smooth}$. Between the smoothing operations, residuals are calculated ($\mathcal{O}_{\rm residual}$), restricted ($\mathcal{O}_{\rm restrict}$), and interpolated from the coarser solution ($\mathcal{O}_{\rm interpol}$) after a recursive call in the V-cycle. Similar to FMM, we can count the total time accumulated over all MPI ranks, 
\begin{eqnarray}
T_{\rm MG} &\approx N\Bigl[ N_{\rm cycle}
\Bigl(
N_{\rm smooth}\,\mathcal{O}_{\rm GS} +
\mathcal{O}_{\rm residual + restrict + interpol}
\Bigr)\Bigr].
\end{eqnarray}
Again, the operators should have a level-dependence, but is dropped here for brevity, and we include the main contributions. Counting the FLOPs yields the following: $\mathcal{O}_{\rm GS}=9$, $\mathcal{O}_{\rm residual}=10$, $\mathcal{O}_{\rm restriction}=1$, and $\mathcal{O}_{\rm interpol}=16$. We thus find:
\begin{eqnarray}
T_{\rm MG} \approx 27 N N_{\rm cycle} +  9 N N_{\rm cycle} N_{\rm smooth}.
\end{eqnarray}
Despite its dependence on $N_{\rm cycle}$ and $N_{\rm smooth}$, the MG entails a significantly smaller number of FLOPs than the FMM solver, roughly 30 times smaller. However, scaling tests show that FMM-1 benefits more from increased parallelism than MG. This is because the MG solver is subject to frequent MPI communication, as it requires multiple V-cycles to converge, whereas FMM is executed in a single hierarchical pass. Interestingly, the two algorithms exhibit similar compute times on a single core, even though FMM's FLOPs are much larger. This indicates that both solvers operate in a regime where data movement, rather than FLOP count, dominates performance. Moreover, this is consistent with FMM having a higher arithmetic intensity and more effective memory reuse. As a result, differences in communication and synchronization overheads largely determine their parallel scaling behavior.\footnote{However, as seen in the fully optimized version of MG, more aggressive memory usage to avoid frequent MPI communication overhead and vectorization can also significantly improve MG performance. Therefore, the exact scaling results and speed-ups should be taken with a grain of salt as they will change with not only the degree of optimization, but solver-specific optimization strategies, and target hardware. Further changes to the FMM algorithm to enable AMR and adaptive time stepping will alter the scaling behaviors.}

\section{Conclusions}
\label{sec:conclusion}
In this work, we have described our implementation of the FMM in RAMSES, using a unigrid configuration and isolated boundary conditions. Like the multigrid solver, FMM is a scalable method that solves the Poisson equation with computational cost $\mathcal{O}(N)$. We first construct the FMM hierarchy on top of the computational grid, introducing a level offset of $\Delta \ell$. The FMM algorithm is then executed in a single upward and downward sweep. During the upward pass, we compute multipole moments up to quadrupole order and evaluate them at the centers of the FMM cells on each level by shifting and accumulating contributions from finer levels. We subsequently decompose the force field recursively into far-, intermediate-, and near-field components. The far-field at a given level is identified with the intermediate-field from the coarser level, while the near-field is treated as the intermediate-field on the next finer level. The well-separated cells that form the intermediate-field constitute the interaction list, which we process by transforming multipoles into local expansions up to third order (M2L). These local expansions are then shifted and propagated downward (L2L) to provide the far-field contribution on the next finer level. Finally, the gravitational field at the centers of the computational cells is obtained by applying L2P for the far-field, M2P for the intermediate-field, and a P2P (direct summation) for the near-field. Our implementation achieves computational speedups by precomputing the interaction list and evaluating multipole moments about cell centers, thereby reusing the fixed cell geometry throughout the calculation.

Our unigrid tests indicate that FMM achieves accuracy comparable to MG for smooth density profiles and is particularly beneficial for astrophysical problems with isolated boundary conditions, because MG produces larger errors near the boundaries when approximate Dirichlet conditions are imposed. In contrast, for sharp profiles, such as the single-charge case, we find that the errors are larger for FMM, and the method exhibits box-shaped error patterns caused by the discontinuity of the local expansion at the boundaries of the FMM grids. Interestingly, our FMM algorithm achieves very similar errors with only a small number of multipoles $n=2$ and a Taylor expansion order of $p=3$. Furthermore, the force accuracy is comparable for $\Delta \ell = 1$ (FMM-1) and $\Delta \ell = 2$ (FMM-2), with only marginal improvement in the latter, which replaces the final intermediate-field evaluation with direct summation.

Finally, we performed strong and weak scaling studies for FMM and MG. We find that FMM-1 is faster than MG using a similar framework and level of optimization. It exhibits however better scalability overall. We also find that FMM-1 scales consistently more efficiently than FMM-2, since it relies more heavily on higher–arithmetic-intensity operations---specifically intermediate-field interactions, rather than primarily on direct summation. Therefore, given that the difference in solution quality is marginal between FMM-1 and FMM-2, FMM-1 is always favorable. More importantly, our findings reveal that the overall FLOP count for FMM is roughly 30 times higher than that for MG. However, operations in FMM have significantly higher arithmetic intensity than operations in MG---Gauss–Seidel, residual computation, restriction, and interpolation---leading to speedups and better scaling behaviors with FMM. Moreover, MG requires several V-cycles to converge, resulting in more frequent MPI communication, which impacts scalability.

Our next paper will focus on FMM in AMR. Previous studies and implementations involving FMM have focused on non-adaptive schemes. Moreover, although FMM has been modified and tested in AMR schemes \citep{Hrycak_Rokhlin1998,Cheng_Greengard_Rokhlin_1999,Ethridge&Greengrad2001}, to our knowledge, it has not been fully treated in an adaptive time stepping scheme. However, adaptive time stepping is crucial in accurately solving the evolution of dynamical systems of our interest and needs to be addressed. 

Our new algorithm in preparation will involve two main modifications. First, different levels of refined cells will have unique FMM hierarchies (or FMM trees) to enable adaptive time stepping. Thus, each tree will only collect multipoles from the leaf cells, and the local expansions will be evaluated from multipoles assigned to different trees. This construction of multiple FMM trees enables adaptive time stepping, where only finer levels become active, while coarser levels are frozen in a previous coarse time-step. Secondly, we will define an additional field, namely the \textit{nearest field}. Compared to the near field, which refers to the neighborhood of the target FMM cell, the nearest field is a collection of direct neighbors of the target AMR cell. The nearest field at level $\ell$ will become the near field at level $l+1$. When evaluating the direct force at level $\ell$, we check if the cells in the near field are further refined, and evaluate direct force from the finer cells. This will enforce the symmetry of forces between direct force evaluated at level $\ell$ and level $\ell+1$.

We note that the scalability of FMM in AMR could be affected due to the modified data structure and algorithm. Especially, the new algorithm necessitates iteration over the existing trees to aggregate force contribution from leaf cells of different refinement levels, and not all of the cells in the interaction list will be available as the FMM tree is built on patches rather than on the full domain as in the unigrid case.

We also emphasize that our FMM algorithm can be further improved in both accuracy and scalability. In particular, characteristic boxy error patterns can be mitigated by employing higher-order multipoles. Alternatively, these errors can be reduced by applying random affine transformations to the computational domain, allowing them to decorrelate and partially cancel, thereby better preserving global physical properties \citep{Barnes2026}. On the performance side, scalability can be enhanced through more aggressive caching and paging strategies to reduce the frequency of MPI communication. Finally, attributed to the nature of FMM involving compute-intensive operations, the algorithm is well-suited to data-parallel acceleration. This will enable efficient vectorization on CPUs as well as acceleration on GPUs, a direction we plan to pursue in future work.

\section*{Acknowledgements}

J.-Y.L thanks Nick Gnedin for insightful discussion. This material is based upon work supported by the National Science Foundation (NSF) and the
U.S.-Israel Binational Science Foundation (BSF) under Award Number 2406558 and Award Title ``The Origin of the Excess of Bright
Galaxies at Cosmic Dawn''. The authors are also pleased to acknowledge that the work reported in this paper was performed substantially using Princeton University’s Research Computing resources, specifically the Stellar cluster.

\section*{Data Availability}

The FMM implementation in the prototype code {\ttfamily mini-ramses} is available from the authors upon reasonable request.



\bibliographystyle{mnras}
\bibliography{example} 




\appendix

\section{The Effect of Level Difference in FMM }
\label{sec:fmm_diff}
In this section, we demonstrate the effect of the level difference between the actual cells in the computational domain and the finest FMM cells, or $\Delta \ell$. Increasing the level difference will generally result increasing accuracy as larger regions will be calculated by direct summation, rather than from multipole expansion terms at the expense of increased computational cost (see Section~\ref{sec:scaling} for more information).

\begin{figure*}
\includegraphics[width=\textwidth]{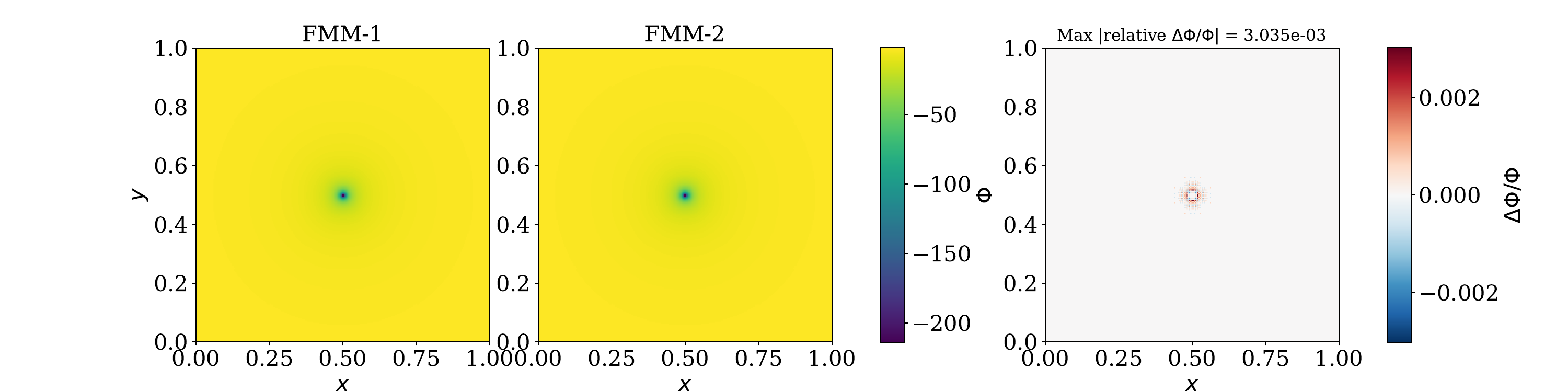}
    \caption{Comparison of potential maps along a slice in the plane containing a single point charge with mass $GM = 1$ for FMM-1 (\textit{left}) and FMM-2 (\textit{center}), with their relative error shown in the \textit{rightmost panel}. The computational domain has a box size of 1 and is resolved on a $256^3$ grid.}
    \label{fig:fmm_diff}
\end{figure*}

In Fig.~\ref{fig:fmm_diff}, we compare the potential maps between FMM-1 ($\Delta \ell=1$) and FMM-2 ($\Delta \ell=2$) for a single charge at the center, as done in Fig.~\ref{fig:single}. The differences are miniscule and we find that a relative difference is less than 0.3\%. Since the actual density distributions will be a convolution of this ideal single-charge configuration, with small and symmetric errors, the actual gain in accuracy from increasing $\Delta \ell \ge 2$, or performing a more direct summation, would be negligible. Considering that FMM-1 scales better than FMM-2, we opt to fix $\Delta \ell = 1$ in our future code design for AMR. 

\section{Test Results for an NFW Halo}
\begin{figure*}
	\includegraphics[width=\textwidth]{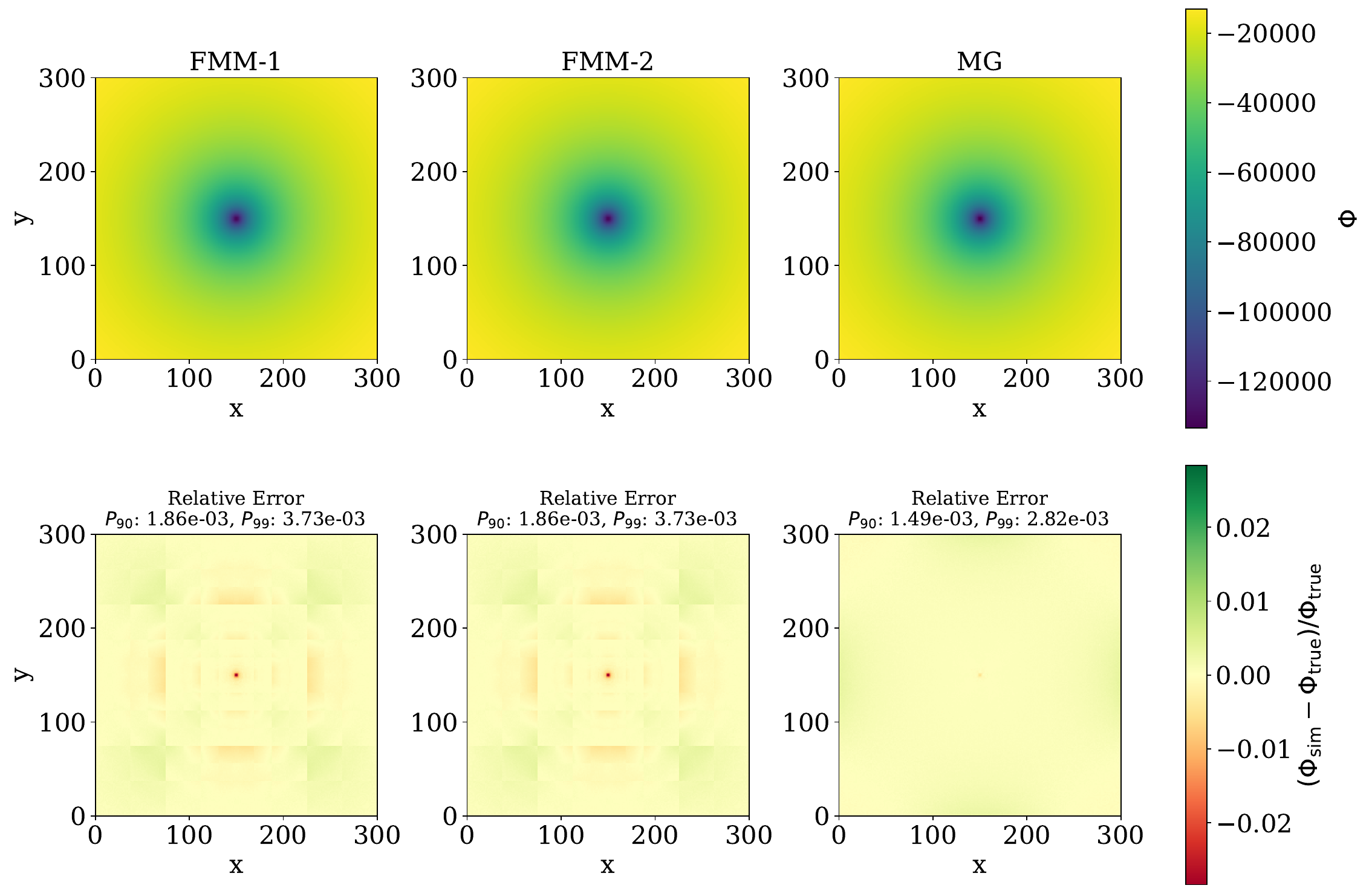} \caption{Comparison of potential maps along a slice through the center of an NFW halo for FMM-1 (\textit{left}) and FMM-2 (\textit{center}), with their difference shown in the \textit{rightmost panel}. The NFW halo is resolved on a $256^3$ grid, and the values shown are in arbitrary code units.}
    \label{fig:nfw}
\end{figure*}

\begin{figure}
	\includegraphics[width=\columnwidth]{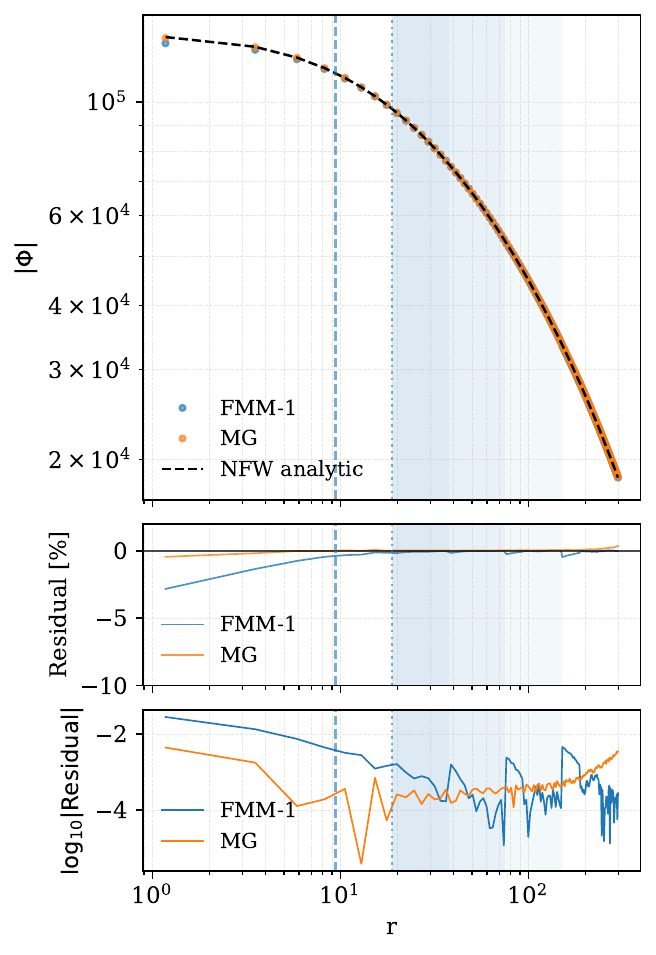}
    \caption{Gravitational potential (\textit{top}) for an NFW halo presented in Fig. \ref{fig:nfw}. Results from FMM-1 (\textit{blue}), MG (\textit{orange}), and the analytic solution (\textit{black dashed line}) are shown. The \textit{middle panel} shows the residual, while the bottom panel shows $\log_{10}$ of the residual. Similar to Fig.~\ref{fig:single_err} the $+x$ direction is depicted, while fixing $y$ and $z$ to slice through the middle of the computational domain. Although the NFW halo exhibits an extended mass profile rather than a single-point charge configuration, we still depict similar force indicators under the assumption that the dominant contribution originates from the central region. The vertical lines are the same as Fig.~\ref{fig:single_err}.}
    \label{fig:nfw_err}
\end{figure}

\label{sec:nfw}
In this section, we show the test result for an NFW halo. In Fig.~\ref{fig:nfw}, we compare the potential maps obtained from FMM-1, FMM-2, and MG. Relative errors are also shown, where the ground truth is derived from a $1/r^2$ direct force calculation of dark matter particles onto the cell centers. The mean relative errors exhibit similar levels across all solvers, and a distinct error arises at the center where the profile peaks for FMM, due to its susceptibility to errors from the mass deposition onto the grid, as we observed for the single-charge test case. Fig.~\ref{fig:nfw_err} demonstrates that FMM indeed suffers from discontinuities of local expansions at the cell boundaries of FMM, while being also more robust than MG closer to the boundaries.

\begin{figure}
    \includegraphics[width=\columnwidth]{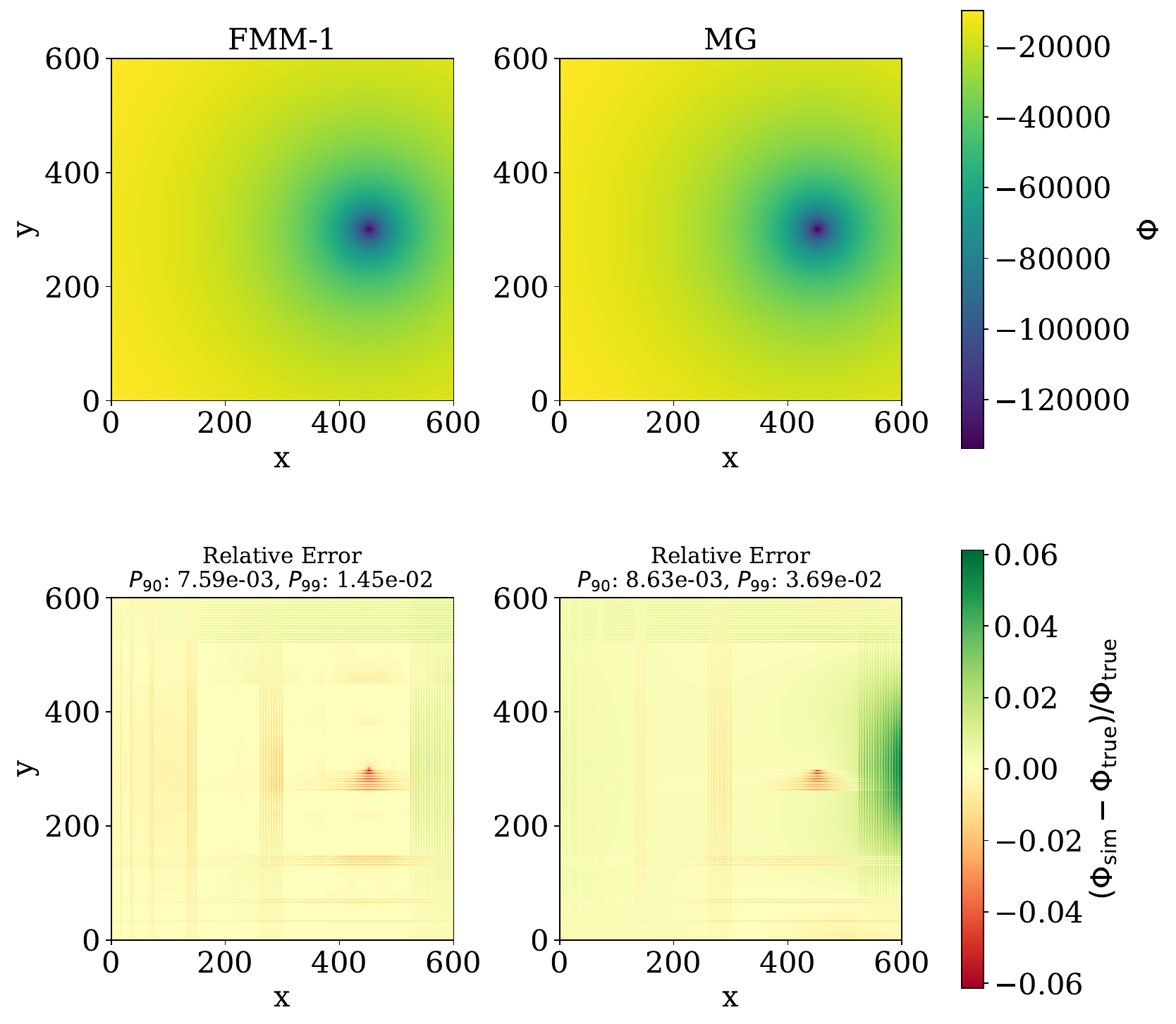}
    \caption{Same as Fig.~\ref{fig:nfw}, but with the center of the NFW halo displaced in the $+x$ direction by $L_{\rm box}/4+\Delta x/2$, where $L_{\rm box}=600$ and $\Delta x=L_{\rm box}/256$. For brevity, we show the results from FMM-1 (\textit{left panels}) and MG (\textit{right panels}).}
    \label{fig:shifted_nfw}
\end{figure}

We also test the asymmetric, non-ideal case, where the NFW halo is displaced from the center. In Fig.~\ref{fig:shifted_nfw}, the particles are shifted to the $+x$ direction by $L_{\rm box}/4+\Delta x/2$, where $L_{\rm box}=600$ and $\Delta x=L_{\rm box}/256$. The latter amount of $\Delta x/2$ effectively shifts the peak to be located at the center of the edge of the cells (shared by four cells) instead of the corner (shared by eight cells). We observe the same boxy errors of FMM and bias of MG towards the boundaries overlaid on top of the shared error patterns that are attributed to the error in the mass deposition onto the grid. While the accuracy of both FMM and MG solutions degrades with respect to the ideal case, we observe that MG is more susceptible to the shift. This is due to the fact that, as the peak moves closer to the right boundary, the bias introduced by the inexact Dirichlet boundary condition becomes more pronounced.

\section{Errors on the Forces}
Despite Poisson's equation being the master equation of our solvers, it is the gravitational force that actually acts on the particles or fluid elements. In this section, we report the errors in the forces. In RAMSES, the force is calculated using a 5-point finite-difference stencil. For particles, the forces are interpolated according to the mass deposition scheme.

\begin{figure}
    \includegraphics[width=\columnwidth]{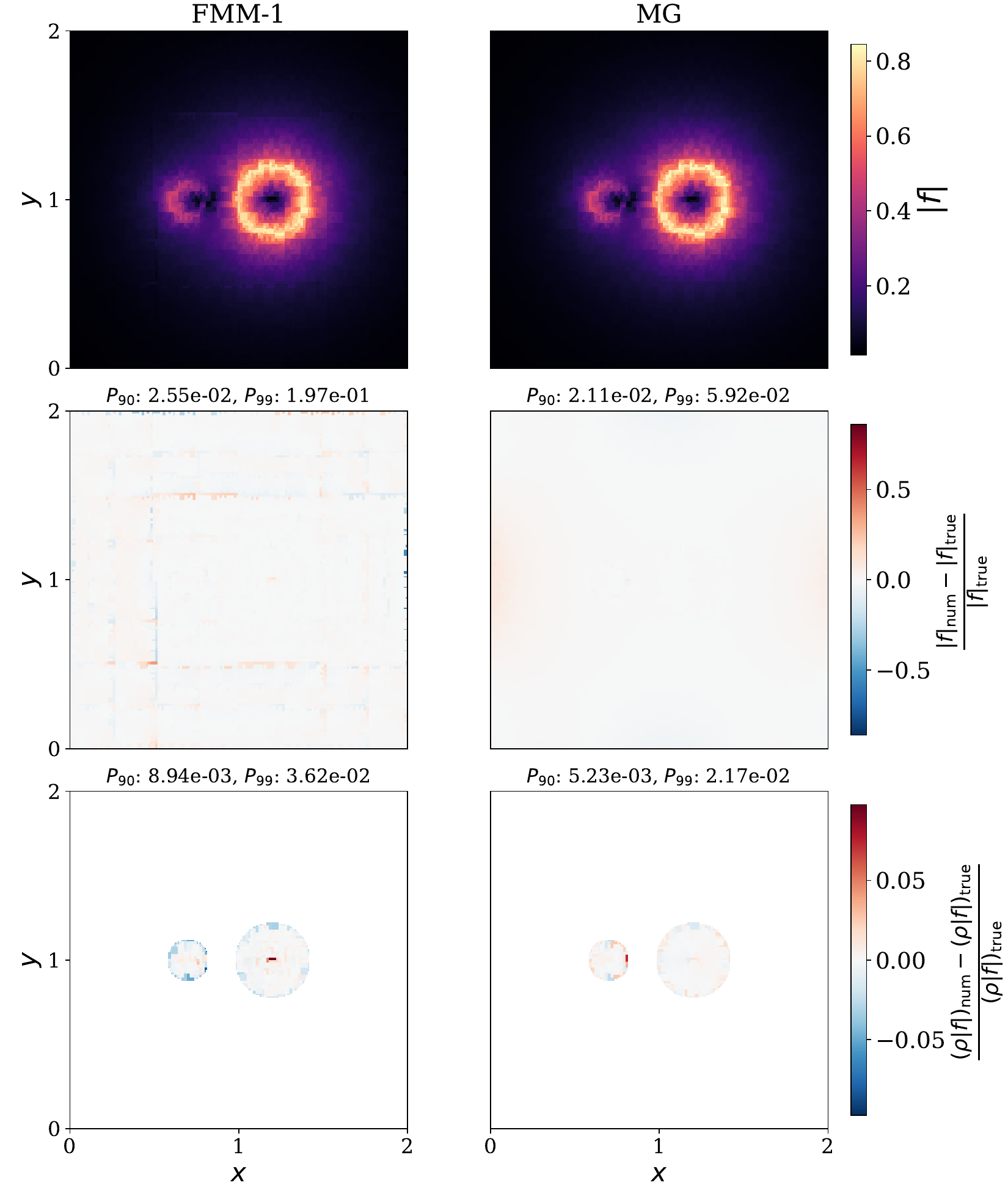}
    \caption{Error maps of the force magnitudes ($|f|$) for the spheres test case previously shown in Fig.~\ref{fig:spheres}. For brevity, we only show results for FMM-1 (\textit{left panels}) and MG (\textit{right panels}). The \textit{top row} presents the force magnitude, the \textit{middle row} the relative error, and the \textit{bottom row} the relative error weighted by the density ($\rho=1$ inside the spheres). Although FMM and MG respectively exhibit boxy error patterns and boundary-biased errors over the full domain, the subdomain where the density is non-zero exhibits smaller errors of comparable scale, with 90th-percentile errors below one percent.}
    \label{fig:spheres_force}
\end{figure}

Fig.~\ref{fig:spheres_force} presents the force magnitude for the double uniform spheres test case shown in Fig.~\ref{fig:spheres}. From the full domain, we observed similar boxy error patterns for FMM as the gradients pick up the boxy patterns from the potential field map. The MG solutions still exhibit a bias toward the box boundaries, although it is less pronounced due to the smoothness of the bias. The 90th-percentile error shows similar levels, while the 99th-percentile becomes slightly worse, especially due to the box edge of FMM now being susceptible to the Dirichlet boundary conditions for the ghost cells. However, once limited to the domain where the density field is non-zero, where the force is truly effective, the forces are fairly accurate with less than a percent deviation from the ground truth for the 90th-percentile.

\bsp	
\label{lastpage}
\end{document}